\documentclass[12pt,notoc,nohyper]{JHEP}

\usepackage{epsfig}
\usepackage{amsmath}
\usepackage{amsfonts}

\title{The exact renormalisation group equation and 
the perturbed unitary minimal models}
\author{Lars Kj\ae rgaard \\ 
Department of Mathematical Sciences \\
University of Durham, Durham DH1 3LE, U.K. \\
\tt{lars.kjaergaard@durham.ac.uk}}

\abstract{The exact renormalisation
group equation is studied for a two dimensional theory with
exponential interaction and a background charge at infinity. The
motivation for studying this interaction is the flow between 
unitary minimal models perturbed by $\Phi_{(1,3)}$, and their
realisation in terms of a quantum group restricted
sine-Gordon model.}

\preprint{DTP/00/44}

\begin{document}

\section{Introduction}
To get a better understanding of the space of quantum field 
theories, it 
is important to study the renormalisation group flow that determines
its structure.
 Two dimensional quantum field theories serve as a good starting
point for two reasons. Firstly, the fixed 
points of the renormalisation group are two dimensional conformal field 
theories which are well understood due to their large symmetry, and secondly,
Zamolodchikov's $c$-theorem states that for a unitary theory
the renormalisation group flows are irreversible flows in the
coupling constant space. 


The exact renormalisation group equation is a functional differential 
equation which describes how the 
wilsonian effective action must vary when the cut-off changes so
that the physics is invariant. It is derived by integrating
out the ultra-violet degrees of freedom in the partition function.
In this paper the exact renormalisation group equation is studied for
an exponential interaction and a background charge at infinity.

The motivation for considering this case is that 
the renormalisation group is known to flow between certain unitary 
conformal field theories, namely the unitary minimal models. 
A two dimensional conformal field theory is specified by its field
content, the operator product coefficients and 
the value of the Virasoro central charge $c$ 
that measures the number of massless degrees of freedom in the theory. 
The unitary minimal models ${\cal M}_m$ ($m=3,4,\ldots$) are unitary conformal 
field theories with a finite number of primary fields (the highest weight 
vectors 
of the Virasoro algebra) and a central charge $0<c=1-\tfrac{6}{m(m+1)}<1$.

The unitary minimal models ${\cal M}_m$ have the same ($m-1$) critical
behaviour as Landau-Ginzburg theory with a bosonic field and even polynomial
interactions up to the power $2(m-1)$ \cite{zamsjnp86}.
The minimal model ${\cal M}_m$ perturbed by the relevant field
$\Phi_{(1,3)}$ was in \cite{alzam} argued 
to have a renormalisation group flow from  ${\cal M}_m$ in the
ultra-violet to 
${\cal M}_{m-1}$ in the infra-red limit using the thermodynamic Bethe
ansatz. 
Indeed, it was shown
perturbatively 
in the double scaling
limit of small coupling $g$ and large $m$ that the infra-red
central charge $c_{IR}$ satisfies that 
$c_{IR}=c_{{\cal M}_{m-1}}+
O(m^{-4})$. 
This calculation was performed to first order in 
\cite{zamorig,cardyludwig} and second order in \cite{laessig}.
 The ultra-violet fixed
point is at $g_{UV}=0$ and in \cite{zamorig,cardyludwig} it was shown
that the infra-red fixed point is at $g_{IR}\propto \tfrac{4}{m+1}$ so
that perturbation theory is a good approximation for the infra-red
theory in the limit $m\rightarrow \infty$ where $g_{IR}\rightarrow
0$.


The exact renormalisation group equation is not solvable and has to be
approximated, 
we will do this by including only an  exponential
interaction together with a background charge in the wilsonian
effective action.   
Using 
the equivalence between the perturbed minimal models and 
the quantum group restricted sine-Gordon model,
we will argue that this interaction describes the renormalisation group
flow of the unitary minimal models perturbed by $\Phi_{(1,3)}$. 

In section 2 the unitary minimal model perturbed by $\Phi_{(1,3)}$
is described together with its realisation as a quantum group
restricted sine-Gordon model. In section 3 the exact renormalisation
group equation 
is introduced, and it is approximated by only allowing relevant
operators in the effective action, which for the minimal models are
given by exponential operators. The wilsonian effective potential does
not contain any field derivatives so the approximation is similar to
the local potential approximation. The non linear term in the exact
renormalisation group equation is approximated using the operator
product expansion. 

For the perturbed minimal models a renormalisation
group equation is obtained where all higher order terms in the
coupling are contained
in the off-critical structure constant for the operator product
expansion. This renormalisation group equation is valid for all $m>3$,
and in the perturbative limit $m\rightarrow \infty$ it is equal to the
perturbative renormalisation group equation.



\section{Unitary minimal models perturbed by $\Phi_{(1,3)}$}
We denote by ${\cal M}_m$ the unitary minimal model that 
has the spin zero 
primary fields $\Phi_{(p,q)}$, $1\le p \le m-1$, $1\le q\le m$, 
 and a  central charge $c=1-\tfrac{6}{m(m+1)}$. 
$\Phi_{(p,q)}$ has the conformal scaling dimension 
$h_{(p,q)}=\bar{h}_{(p,q)}=\tfrac{((m+1)p-mq)^2-1}{4m(m+1)}$ which
shows that 
$\Phi_{(p,q)}=\Phi_{(m-p,m+1-q)}$ leaving only $\tfrac{m(m-1)}{2}$ different primary fields.  
The unitary minimal model ${\cal M}_m$ perturbed by the relevant operator
$\Phi_{(1,3)}$ can formally be written as\footnote{The standard conventions are
used (see \cite{zamorig}), with the   
normalisation $d^2z\equiv 2idz \wedge d\bar{z}=dx\wedge dt_E =d^2x$,
and positive structure constant $C^{(1,3)}_{(1,3)(1,3)}=C_{\Phi\Phi}^\Phi>0$.} 
\begin{equation}
  \label{minmod}
  S={\cal M}_m+\lambda \int d^2 z \ \Phi_{(1,3)}(z,\bar{z}),
\end{equation}
and the theory is defined by 
the correlators
\begin{equation}
  \label{partition}
  \langle {\cal O}_1(x_1) \cdots {\cal O}_n(x_n)\rangle = 
\frac{\langle {\cal O}_1(x_1)\cdots {\cal O}_n(x_n)e^{-\lambda \int d^2z
\Phi_{(1,3)}}\rangle_{{\cal M}_m}}{
\langle e^{-\lambda \int d^2z \Phi_{(1,3)}} \rangle_{{\cal M}_m}}
\end{equation}
where $\langle \cdots \rangle_{{\cal M}_m}$ is the correlator in the
minimal model ${\cal M}_m$ and 
${\cal O}_i(x)$ are the local scaling fields in the theory. 
The renormalisation group eigenvalue for $\Phi_{(1,3)}$ follows 
from \eqref{minmod} using dimensional analysis:  
$y=2-\Delta=\tfrac{4}{m+1}$ where $\Delta=2h_{(1,3)}$. 
$\Phi_{(1,3)}$ is therefore a relevant operator ($y>0$), and 
from the form of $h_{(p,q)}$ above it
directly follows
that the other relevant primary fields are 
$\Phi_{(p,p+s)}$ with  $-1\le s\le 2$ and $0< p+s<m+1$ for $1\le p\le m-1$.
There are therefore $2m-3$ different relevant primary fields including 
$\Phi_{(1,1)}=\boldsymbol{1}$. 
These are the only relevant scalar fields as all the spin zero
descendants will have scaling dimension $2h_{(p,q)}+2n$ with $n\in \mathbb{N}$.

In the Coulomb gas \cite{dotsenko} 
(or Feigin--Fuchs, Dotsenko--Fateev) representation 
the minimal model ${\cal M}_m$ is realised as a
Lagrangian free field theory with a background charge $-2\alpha_0$ at 
infinity and a central charge $c=1-24\alpha_0^2$. $\alpha_0>0$ so that
$\alpha_0=1/\sqrt{4m(m+1)}$.  
The action is then  
\begin{equation}
  \label{actionSg}
  S=\frac{1}{8\pi}\int d^2x \sqrt{g(x)}(g^{\mu \nu}(x)
\partial_\mu \phi (x)\partial_\nu \phi (x)+2\sqrt{2}i \alpha_0 \phi (x)R(x))
\end{equation}
on the Riemann sphere $\mathbb{C}\cup \{ \infty \}$. The primary fields
are the vertex operators
$V_{\alpha_{r,s}}(x)=e^{i\sqrt{2}\alpha_{r,s}\phi(x)}$ 
with
$\alpha_{r,s}=\tfrac{1-r}{2}\alpha_+ +\tfrac{1-s}{2}\alpha_-$ where
$\alpha_- \alpha_+=-1$ and
$\alpha_-+\alpha_+=2\alpha_0$ so that
$\alpha_-=-\sqrt{\tfrac{m}{m+1}}$. 
The vertex operator conformal dimension becomes 
\begin{equation}
\label{nyconfdim}
h_{r,s}=\bar{h}_{r,s}=\alpha_{r,s}(\alpha_{r,s}-2\alpha_0)
\end{equation} 
and therefore the primary field 
$\Phi_{(1,3)}$ is represented as
$e^{-i\sqrt{2}\alpha_-\phi}$.
Note that
$h_{\alpha_{\pm}}=\alpha_{\pm}(\alpha_{\pm}-2\alpha_0)=1$. 
To calculate correlators of vertex operators screening charges 
$Q_\pm =\int d^2z V_{\alpha_\pm}$ are
added so that the overall charge of the correlator vanishes \cite{dotsenko}. 
The presence of a background charge changes the 
energy-momentum tensor
\begin{equation}
  \label{energy}
T=  -\frac{1}{2}\partial \phi \partial \phi \xrightarrow{-2\alpha_0} 
T=-\frac{1}{2}\partial \phi \partial \phi+
i\sqrt{2}\alpha_0\partial^2 \phi,
\end{equation}
and therefore also 
the scaling behaviour of the theory lowering the central charge
from the value of the free theory $c=1$ to $c=1-24\alpha_0^2$. 

In \cite{felderleclair,smirnov,thpm} it was observed that the
perturbed unitary minimal model \eqref{minmod} is equivalent
to a quantum group restriction of the sine-Gordon theory. 
The sine-Gordon  theory is given by the euclidean action 
\begin{equation}
  \label{SG}
  S_{SG}=\frac{1}{4\pi}\int d^2x \left(\frac{1}{2}
\partial_\mu \phi(x)\partial^\mu
  \phi(x)-4\pi \tilde{\lambda} \cos (\beta
  \phi(x)/\sqrt{4\pi})\right). 
\end{equation}
In two dimensions all polynomial interactions are super-renormalisable and
all divergences appear in the tadpole diagrams which are 
the self contractions of a vertex.
The renormalisation can be done by normal ordering which removes all self 
contractions, 
and this corresponds to a renormalisation of $\tilde{\lambda}$,
but $\beta$ and $\phi$ are not renormalised \cite{coleman}.

In \cite{felderleclair} the quantum group restriction of the
sine-Gordon model is performed by adding a background charge
$-2\alpha_0$ at infinity.
The background charge will again change the
energy-momentum tensor as in \eqref{energy} and the two vertex
operators in the potential will now scale differently.
Two different couplings are needed, we will write this as
\begin{equation}
  \label{changebackg}
  \tilde{\lambda} \cos (\beta \phi/\sqrt{4\pi})\xrightarrow{-2\alpha_0}
  \frac{\tilde{\lambda}}{2} 
e^{-i \beta  \phi/\sqrt{4\pi}}+\frac{\tilde{\lambda}_-}{2} 
e^{i\beta\phi/\sqrt{4\pi}}.
\end{equation}
Analogous to the Coulomb gas representation for the minimal models the
background charge $\alpha_0$ is determined from the central charge
$c=1-24\alpha_0^2=1-\tfrac{6}{m(m+1)}$.  
$\beta$ is chosen from the requirement that
$e^{i\beta \phi/\sqrt{4\pi}}$ becomes marginal (i.e.\ $h=\bar{h}=1$)
so that it survives in the ultra-violet limit.
$\beta=\sqrt{8\pi}\alpha_-=-\sqrt{8\pi}\sqrt{m/(m+1)}$. 
The conformal dimension of
$e^{-i\beta\phi/\sqrt{4\pi}}$ is then
$h=\tfrac{m}{m+1}-\tfrac{1}{m+1}=\tfrac{m-1}{m+1}=h_{(1,3)}$ 
and   $e^{-i\beta \phi/\sqrt{4\pi}}$ represents the
perturbing operator $\Phi_{(1,3)}$
\footnote{The ultra-violet limit of the restricted sine-Gordon model is a
Liouville theory which is equivalent to the unitary minimal models
${\cal M}_m$, the perturbed minimal models can therefore also be seen
as a perturbed Liouville theory \cite{liouvrefs}.}.
Only relevant operators are included in the effective action, as
explained below, so the marginal screening term $e^{i\beta\phi
/\sqrt{4\pi}}$ (which is needed when considering correlators) is 
excluded.

In \cite{felderleclair} it was argued that the perturbed unitary minimal model
\eqref{minmod} is a quantum group restriction of the
sine-Gordon theory, i.e.\ a massive theory, with the coupling taken as
$\tilde{\lambda}=-\lambda$ in \eqref{SG}. 
Here we are only interrested in the massless flow  
to ${\cal M}_{m-1}$ ($\lambda>0$ in
\cite{zamorig}) 
where the limit $m\rightarrow
\infty$ is given by perturbation theory. We will therefore take
the opposite sign, $\tilde{\lambda}=k \lambda$ for some $k\in
\mathbb{R}_+$, and in  the exact renormalisation group equation
we consider the action 
\begin{equation}
 \label{SG2}
  S=\frac{1}{4\pi}\int d^2x \left(\frac{1}{2}
\partial_\mu \phi(x)\partial^\mu
  \phi(x)-4\pi \frac{\tilde{\lambda}}{2} e^{-i\beta
  \phi(x)/\sqrt{4\pi}}
+\sqrt{2}i \alpha_0 \phi (x)R(x)\right)
\end{equation}
 as a model for the perturbed unitary minimal model \eqref{minmod}. 
The operator $\Phi_{(1,3)}$ in \eqref{minmod} is normalised 
as in \cite{zam9807} so
that the ultra-violet limit of correlators in the perturbed theory
\eqref{partition} equals the corresponding correlators
in ${\cal M}_m$ where we take $\langle
\Phi_{(1,3)}(1)\Phi_{(1,3)}(0)\rangle_{{\cal M}_m}=1$. 
The scaling dimension of $V_{(1,3)}$ and
$\Phi_{(1,3)}$ are equal\footnote{In \cite{alzamsgmass} 
the exact relation
between the sine-Gordon coupling $\tilde{\lambda}$ (i.e.\ without a
background charge, so that the scaling dimensions of $\cos
(\beta\phi/\sqrt{4\pi})$ and $\Phi_{(1,3)}$ here differ) and $\lambda$
was established $\lambda=
\tfrac{\tilde{\lambda}^2\pi}{4(1-\beta^2/(4\pi))(3\beta^2/(8\pi))}
(\gamma (1-\beta^2/(8\pi))^3\gamma (3\beta^2/(8\pi)))^{1/2}$ 
using the Coulomb gas representation of
minimal models \cite{dotsenko} ($\gamma (x)=\Gamma(x)/\Gamma(1-x)$). 
We do not use this representation here
because we want the ultra-violet limit to be ${\cal M}_m$, and not as
in 
the sine-Gordon model where 
it is the free theory with $c=1$, also we want a
linear relation between $\lambda$ and $\tilde{\lambda}$.}
so the normalisation constant relating them is independent of
$\tilde{\lambda}$, and is given by the formulas in 
\cite{dotsenko}\footnote{${N_{(1,3)}}^2=2\pi^2\tfrac{\gamma (1-\alpha_-^2)^3
\gamma (3\alpha_-^2-1)}{(1-2\alpha_-^2)(2\alpha_-^2-1)}$ with
$N_{(1,3)}>0$.}  $V_{(1,3)}=N_{(1,3)}\Phi_{(1,3)}$, we
therefore take
$\tilde{\lambda}={N_{(1,3)}}^{-1}\lambda$.

For a non zero background charge  the scaling of the theory  
is not determined by the usual 
$\boldsymbol\beta$-function related to the ultra-violet
divergences, but by a generalised $\boldsymbol{\beta}$-function as 
discussed in \cite{grisaru}. The generalised $\boldsymbol{\beta}$-function
takes into account the change in the energy-momentum tensor 
when the background charge is added. 
To lowest order the background charge adds a term \cite{grisaru}
\begin{equation}
  \label{addterm}
  \boldsymbol{\beta}(\tilde{\lambda})=\Lambda \frac{\partial
  \tilde{\lambda}}{\partial \Lambda}
\rightarrow \tilde{\boldsymbol{\beta}}(\tilde{\lambda})
= \boldsymbol{\beta}(\tilde{\lambda})-2\cdot 2\alpha_{(1,3)} \alpha_0 
\tilde{\lambda}
\end{equation}
to the normal perturbative $\boldsymbol{\beta}$-function for the
interaction term $\tilde{\lambda}
e^{i\sqrt{2}\alpha_{(1,3)} \phi(x)}$ in the action. This follows from the
change in the scaling dimension of $\Phi_{(1,3)}$: $\Delta \rightarrow \Delta
-2\cdot 2\alpha_{(1,3)}\alpha_0$  with a background
charge \eqref{nyconfdim}\footnote{The $\boldsymbol{\beta}$-function starts as
$-y \tilde{\lambda}$ where $y$ is the renormalisation group
eigenvalue.}. We  discuss below how this term appears in the exact 
renormalisation group. 

\section{The exact renormalisation group}
The wilsonian effective action at the scale $\Lambda$ is obtained by
`integrating out' momentum modes between $\Lambda$ and $\Lambda_0$,
where $\Lambda_0$ is the fundamental wilsonian cut-off. The exact
renormalisation group equation\footnote{Or the Wilson-Polchinski 
renormalisation group equation \cite{wilson,polchinski}.} 
then describes how the effective action
must change so as to describe the same physics when $\Lambda$ changes. 
Following \cite{polchinski}, the partition function is written as 
\begin{equation}
\label{partitionP}
Z(\Lambda)=\int D\phi\  e^{-S[\phi]}=
\int D\phi\  e^{-\frac{1}{2}\int 
\frac{d^2p}{(2\pi)^2}\phi(p)\phi(-p)p^2 K^{-1}(p^2/\Lambda^2)-V(\phi,\Lambda)}
\end{equation}
in a continuum formulation with the cut-off propagator 
$\tfrac{K(p^2/\Lambda^2)}{p^2}$. $K(p^2/\Lambda^2)$  is constant for small 
$s=p^2/\Lambda^2$ and vanishes faster than any power for large $s$. 
The physics must be independent in the choice of
effective scale $\Lambda$ so that
$\Lambda\tfrac{d}{d\Lambda}Z(\Lambda)=0$, and from \eqref{partitionP}
it follows that
\begin{equation}
\Lambda\frac{d}{d\Lambda}Z(\Lambda)=\langle (-\frac{1}{2}\int
\frac{d^2p}{(2\pi)^2}\phi(p)\phi(-p)p^2\Lambda
\frac{\partial K^{-1}(p^2/\Lambda^2)}{\partial
\Lambda}-\Lambda\frac{\partial V(\phi,\Lambda)}{\partial \Lambda})
\rangle=0.
\end{equation}
This condition holds when the wilsonian effective
potential satisfies the operator 
equation\footnote{A constructive proof is given in 
\cite{zinn-justin} by explicit functional integration.}
\begin{equation}
\label{erg2}
  \Lambda \frac{\partial}{\partial \Lambda} V(\phi,\Lambda)  
 =
\frac{1}{2}\int\! d^dx
   d^dy \int \! \frac{d^d k}{(2\pi)^d}e^{ik(x-y)} \frac{2}{\Lambda^2}
K'(\frac{k^2}{\Lambda^2})
 (\frac{\delta^2 V}{\delta \phi(x)\delta \phi(y)}-\frac{\delta V}{\delta
  \phi(x)}\frac{\delta V}{\delta \phi(y)}),
\end{equation}
because the 
partition function then changes as a total derivative (up to a field
independent term) $\Lambda \frac{d}{d\Lambda}Z=\int
d^2p \int D\phi \frac{\delta }{\delta \phi}\Psi=0$ \cite{polchinski}. 
$K'(k^2/\Lambda^2)=\tfrac{dK(s)}{ds}|_{s=k^2/\Lambda^2}$. 
\eqref{erg2} is the exact renormalisation group equation. 
It is a non linear functional equation in $V(\phi,\Lambda)$ 
which is not directly solvable and approximations must be made, either
by truncating the operator space or by performing a derivative expansion 
\cite{morris}. 

We will  make a truncation in 
the  operator space. Firstly, we will only consider relevant operators, 
this approximation becomes exact in the infra-red ($\Lambda\ll
\Lambda_0$) where 
the effective theory is determined by the relevant (and marginal) couplings. 
Secondly, it is known that the primary fields 
$\{ \Phi_{(1,2p+1)}\}$ with $0\le p \le [\tfrac{m-1}{2}]$ form a 
sub-algebra\footnote{$\Phi_{(1,r)}\Phi_{(1,s)}=\sum_p C_{rs}^p \Phi_{(1,p)}$  
where $p\in \{|r-s|+1,|r-s|+3,\ldots,r+s-1 \}$ in steps of two 
\cite{dotsenko}.} 
in ${\cal M}_m$, and the renormalisation group flow 
 from the minimal
 model  ${\cal M}_m$ is therefore spanned by these 
operators\footnote{The  operator structure at
 the non trivial ultra-violet 
fixed point thereby determines the renormalisation group flow.
 When considering perturbations from operators in a sub-algebra, all 
divergences that arise away from the fixed point will be contained in this 
sub-algebra showing that there is a renormalisation group flow  in the
sub-space of the corresponding couplings.}.

In the sub-algebra $\{ \Phi_{(1,2p+1)}\}$ only
$\Phi_{(1,1)}=\boldsymbol{1}$ and
$\Phi_{(1,3)}$ are relevant 
fields\footnote{The operator $\Phi_{(1,3)}$ does not mix with other
relevant operators off criticality \cite{zamorig,cardylesh}.}. 
The exact renormalisation group is only determined up to a field
independent term, 
so we will only consider perturbations with respect to $\Phi_{(1,3)}$
\footnote{The integrable perturbation with $\Phi_{(1,3)}$ was first
considered in \cite{zamolodchikovint}.}.   
None of the relevant operators $V_{(1,n)}$ contain any derivatives in
$\phi$, so the situation is similar to the local potential
approximation of the exact renormalisation group equation. 
The local potential approximation is the lowest order term in a
derivative  expansion of 
$V=V(\Lambda,\phi,\partial \phi,\ldots)$ \cite{ballmorris}. 
The second order term 
containing derivatives $f(\Lambda)(\partial \phi)^2$ leads to a 
renormalisation of $\phi$ and an anomalous
 dimension $\eta>0$. In our approximation there are no derivatives in
$\phi$
because we only keep relevant operators,  and $\phi$ is  not 
renormalised in the sine-Gordon model.
In this approximation the first term in \eqref{erg2}  can 
be rewritten \cite{ballmorris} as
\begin{equation}
  \label{last}
 \frac{1}{\Lambda^2}\int \frac{d^2k}{(2\pi)^2}K'(k^2/\Lambda^2)\int
  d^2x\frac{\partial^2 {\cal V}}{\partial \phi\partial \phi}(x)=
-\gamma_1\int
  d^2x\frac{\partial^2 {\cal V}}{\partial \phi\partial \phi}(x)
\end{equation}
where ${\cal V}(x)$ is the potential density $V=\int d^2 x{\cal V}(x)$,  
and $\gamma_1=-\int \tfrac{d^2 k}{(2\pi)^2}\ K'(k^2)>0$ as $K(k^2)$ is
decreasing. 
In the local potential 
approximation \cite{tokar,ballmorris,filippov}
the second non local term in \eqref{erg2}
\begin{equation}
\label{nonlocalterm}
-\int d^2 x \int d^2 y \int \tfrac{d^2 k}{(2\pi)^2}K'(k^2/\Lambda^2)
e^{ik(x-y)}\frac{\partial {\cal V}}{\partial \phi (x)}
\frac{\partial {\cal V}}{\partial \phi (y)}
\end{equation}
is approximated by 
$\gamma_2\int d^2x (\tfrac{\partial {\cal V}}{\partial \phi (x)})^2$ where 
$\gamma_2 \in \mathbb{R}_+$ depends on the cut-off function $K(s)$. 
This is a good 
approximation in the ultra-violet limit $\Lambda\rightarrow \infty$ 
\cite{filippov}
 where $\int \tfrac{d^2 k}{(2\pi)^2}K'(k^2/\Lambda^2)e^{ik(x-y)}\rightarrow 
K'(0)\int \tfrac{d^2k}{(2\pi)^2}e^{ik(x-y)}=K'(0)\delta^2(x-y)$. 
We consider the 
operator version of \eqref{erg2} and a different approximation
therefore has to be used 
for the term \eqref{nonlocalterm}  
otherwise divergences will appear in the operator product 
$\tfrac{\partial {\cal V}}{\partial \phi (x)}
\tfrac{\partial {\cal V}}{\partial \phi (x)}$, 
whereas \eqref{erg2} is finite for $\Lambda< \infty$. 
The correct form of the  approximation in the 
operator case is obtained from the operator product expansion. For
the spin zero operators is 
\cite{cptzam}
\begin{equation}
  \label{ope2}
  {\cal O}_i(x){\cal O}_j(y)\sim 
\sum_k |x|^{\Delta_k-
  \Delta_i-\Delta_j}\tilde{C}_{ij}^k(\tilde{\lambda},\Lambda)
{\cal O}_k (y),
\end{equation}
where ${\cal O}_k(y)$ is a complete set of local scaling fields.
The non local term \eqref{nonlocalterm} then takes the form 
\begin{equation}
\int d^2x {\cal O}(x)\int d^2y \int \tfrac{d^2k}{(2\pi)^2}e^{ik(x-y)}
K'(k^2/\Lambda^2)h(|x-y|^2)
\end{equation}
for some $h(|x-y|^2)$ and operator ${\cal O}(x)$. 
We use that $\delta^2(x)=\tfrac{1}{\pi}\delta(|x|^2)$ to write 
the ultra-violet limit $\Lambda\rightarrow \infty$ of the $k$ integral
as  $-\int \tfrac{d^2k}{(2\pi)^2}e^{ik(x-y)}K'(k^2/\Lambda^2)\rightarrow 
\gamma_2 \delta(|x-y|^2)$ where again $\gamma_2\in \mathbb{R}_+$ 
depends on the choice of $K(s)$.
For finite $\Lambda$ 
this expression is approximately  
$\gamma_2\delta(|x-y|^2-a/\Lambda^2)$ where $a\in \mathbb{R}_+$
depends on $K(s)$. The cut-off dependence in $a$ is removed by redefining 
$\Lambda^2 \rightarrow a\Lambda^2$, and the approximation becomes 
 \begin{equation}
  \label{reg}
 -\int \frac{d^2k}{(2\pi)^2}e^{ik(x-y)}K'(k^2/\Lambda^2)=\gamma_2\ 
\delta(|x-y|^2-\Lambda^{-2})
\end{equation}
in \eqref{nonlocalterm}\footnote{Adding a function
$f(|x-y|^2,1/\Lambda^2)$ 
satisfying $f\rightarrow
0$ for $\Lambda \rightarrow \infty$ corresponds to a higher order derivative
expansion in $\phi$ \cite{filippov} 
and it is therefore neglected in the approximation 
considered here where there are no derivatives in $\phi$.}.
If the effective potential is written as $V=\tilde{\lambda}^i \int d^2x {\cal
O}_i(x)$, where $[\tilde{\lambda}^i]=y^i> 0$, then  \eqref{erg2} becomes using 
\eqref{last} and the operator product expansion
\begin{equation}
\label{erg3}
\begin{split}
\frac{\gamma_1}{\gamma_2}\int d^2x \Lambda \frac{d}{d\Lambda}{\cal V}(\phi,\Lambda) = &
-\frac{\gamma_1}{\gamma_2}\int d^2x \frac{\partial^2 {\cal
V}(\phi,\Lambda)}{\partial \phi \partial \phi} \\
& +\frac{\gamma_1}{\gamma_2}\frac{\pi}{\Lambda^2}\int d^2 x
\sum_{i,j,k}\tilde{\lambda}^i\tilde{\lambda}^j 
\Lambda^{{\Delta'}_i+{\Delta'}_j-{\Delta'}_k}
{\tilde{C}}_{ij}^{\prime k}(\tilde{\lambda},\Lambda){{\cal O}'}_k(x). 
\end{split}
\end{equation}
Here $V$ and $\phi$ have been rescaled: 
$V\rightarrow \tfrac{\gamma_1}{\gamma_2}V$, $\phi\rightarrow
\sqrt{\gamma_1}\phi$ and 
${\Delta'}_i$ and ${\tilde{C}}_{ij}^{\prime k}$ are the scaling dimension
and structure
constant for the field ${{\cal O}'}_i=\tfrac{\partial {\cal
O}_i}{\partial \phi}$. If ${\cal O}_i$ is a vertex
operator then ${\Delta'}=\Delta$, ${{\tilde{C}}}_{ij}^{\prime k}\propto
\tilde{C}_{ij}^k$ and
${{\cal O}'}_i\propto {\cal O}_i$. 
It follows from \eqref{erg3} that the
implicit dependence in the choice of cut-off function $K(s)$ contained in 
$\gamma_1$ and $\gamma_2$ drops out \cite{ballmorris}.
 The renormalisation group equation is usually written
in terms of dimensionless couplings
$\tilde{g}^i=\Lambda^{-y_i}\tilde{\lambda}^i$ 
so that
$V=\tilde{g}^i\Lambda^2 \int d^2x\tilde{{\cal O}}_i$ where 
$[\tilde{{\cal O}}_i]=0$. 
\eqref{erg3} then becomes
\begin{equation}
\label{erg5}
\begin{split}
\Lambda^2 \int d^2x  \Lambda \frac{\partial \tilde{g}^i(\Lambda)}{\partial
\Lambda}\tilde{{\cal O}}_i= & -2\Lambda^2 \int d^2x \tilde{g}^i\tilde{{\cal
O}_i}-\Lambda^2\int d^2x \tilde{g}^i\frac{\partial \tilde{{\cal O}_i}}{\partial
\phi \partial \phi}\\ 
& +\pi \Lambda^2 \int d^2x \sum_{i,j,k}\tilde{g}^i\tilde{g}^j
{\tilde{C}}_{ij}^{\prime k} \tilde{{{\cal O}'}_k}.
\end{split}
\end{equation}
For the perturbed minimal models we will
use the action \eqref{SG2}, and for the moment neglect the curvature term.
\eqref{erg5} then becomes 
\begin{equation}
\int d^2x e^{-i\beta\phi/\sqrt{4\pi}}
\left(  \Lambda\frac{\partial \tilde{g}}{\partial
 \Lambda}   +2\tilde{g}(\Lambda)+
\tilde{g}(\Lambda)(\frac{-i\beta}{\sqrt{4\pi}})^2 
+\pi \frac{\tilde{g}(\Lambda)^2}{2}
\tilde{C}^\Phi_{\Phi\Phi}(\tilde{g}(\Lambda),\Lambda)
(\frac{-i\beta}{\sqrt{4\pi}})^2 \right)  
=0,
\nonumber
\end{equation}
up to field independent and irrelevant
terms.
Introducing the dimensionless variable $t=\ln
\tfrac{\Lambda_0}{\Lambda}$ (and
$\tilde{x}=\Lambda x$) the renormalisation group equation for
$\tilde{g}$ can then be written as  
\begin{equation}
\label{rg1}
\frac{\partial \tilde{g}(t)}{\partial
t}=\dot{\tilde{g}}=2\tilde{g}(t)-\frac{2m}{m+1}\tilde{g}(t)-
\pi\frac{m}{m+1}\tilde{g}(t)^2 \tilde{C}_{\Phi\Phi}^\Phi(\tilde{g}(t),t).
\end{equation}
Adding the additional term \eqref{addterm} from the background charge
$2\cdot 2\alpha_{(1,3)}\alpha_0 \tilde{g}(t)=\tfrac{2}{m+1}\tilde{g}(t)$ gives
\begin{equation}
\label{ergfin}
\dot{\tilde{g}}(t)=\frac{4}{m+1}\tilde{g}(t)-\pi 
\frac{m}{m+1}\tilde{C}_{\Phi \Phi}^\Phi(\tilde{g}(t),t)\tilde{g}(t)^2=
y\tilde{g}(t)-
\pi \tilde{C}_{\Phi \Phi}^\Phi(\tilde{g}(t),t) \tilde{g}(t)^2\frac{m}{m+1}.
\end{equation}
The contribution from the background charge
to the scaling behaviour can be incorporated directly into the exact
renormalisation group equation if the curvature term in \eqref{SG2} is
taken into account. 
The action \eqref{SG2} is
defined on the Riemann sphere $\mathbb{C}\cup \{ \infty\}$ where all the
curvature is situated at infinity so that the topological invariant 
$\int d^2 \tilde{x}\sqrt{g(\tilde{x})}R(\tilde{x})=8\pi$ is satisfied.   
 $\mathbb{C}\cup \{ \infty\}$ cannot be covered by a single coordinate
 chart, one chart is needed for the flat space and another in the
 neighbourhood of infinity. We take polar coordinates in flat space $x=(r\sin
 \theta,r\cos \theta)$ and define the contribution from infinity as in 
 \cite{zamogzamliou} by a limit value. We consider the theory on a disk
 $\Gamma$ with radius $r\rightarrow \infty$ so that all curvature is at the
 boundary $\partial \Gamma$. 
From the Gauss-Bonnet theorem
 $\lim_{r\rightarrow \infty}\int_{\partial \Gamma}d\theta \ r R(r)=8\pi$ so
 that  $R(r)=\tfrac{4}{r}$ on $\partial \Gamma$. We define as in
 \cite{zamogzamliou} the contribution from infinity to be
$\phi_\infty =\lim_{r\rightarrow \infty}\tfrac{1}{2\pi r}\int_{\partial
  \Gamma}dl\ \phi=\lim_{r\rightarrow \infty}\tfrac{1}{2\pi}\int_{\partial
  \Gamma} d\theta \
\phi$ and similarly for the vertex operators. The effect of the background term
is seen by evaluating the exact renormalisation group equation at infinity
where the curvature is non-vanishing. Hence we replace $\int d^2x$ in
\eqref{erg5} by
$\lim_{r\rightarrow \infty}\tfrac{1}{2\pi}\int_{\partial
  \Gamma}d\theta$. Only the term $\tfrac{\partial V}{\partial
  \phi}\tfrac{\partial V}{\partial \phi}$ in the exact renormalisation group 
equation \eqref{erg2} will give an additional term proportional to 
$e^{-i\beta\phi/\sqrt{4\pi}}$ 
when the background charge is added. This term becomes 
\begin{equation}
\begin{split}
&  \lim_{r\rightarrow \infty}\frac{-1}{2\pi}\int_{\partial \Gamma}\! d\theta
\int \! d^2y \delta
(|x-y|^2-\Lambda^{-2})\frac{2^2(\sqrt{2})^2(i)^2\alpha_0
  \alpha_{(1,3)}}{8\pi}rR(r) e^{i\sqrt{2}\alpha_{(1,3)}\phi(y)}\tilde{g}(t)
\\ = \ & (2)^3 \alpha_0 \alpha_{(1,3)}\tilde{g}(t)\lim_{r\rightarrow \infty}
\int_{\partial \Gamma} \frac{d\theta }{2\pi} \int \frac{d \theta'}{2\pi}
\int \frac{d(r')^2}{2}\delta ((r')^2-\Lambda^{-2})
e^{i\sqrt{2}\alpha_{(1,3)}\phi(x-y')} \\ 
\approx \  & 2^2\alpha_0\alpha_{(1,3)}\tilde{g}(t)\lim_{r\rightarrow \infty}\int_{\partial
  \Gamma}\frac{d\theta}{2\pi}e^{i\sqrt{2}\alpha_{(1,3)}\phi(x)}
= 2^2\alpha_0\alpha_{(1,3)}\tilde{g}(t)V_{(1,3)}(\infty).
\end{split}
\nonumber
\end{equation}
The second to last equation holds in the limit $r\rightarrow \infty$ where 
$r\gg \Lambda^{-1}$.
When the coefficient of $e^{-i\beta\phi/\sqrt{4\pi}}$ from this term
is added to 
\eqref{rg1} then
 \eqref{ergfin} is again obtained, but it now holds 
to all orders in the coupling $\tilde{g}$. 

This shows how the change in scaling
behaviour due to the background charge is seen in the
exact renormalisation group equation when it is evaluated at a point 
with non zero curvature. Equation \eqref{ergfin} is then 
valid to all orders in perturbation theory 
for the chosen truncation of the operator space.
Hence, in 
\begin{equation}
\label{sidsteeq}
\dot{\tilde{g}}(t)=y\tilde{g}(t)-
\pi \tilde{C}_{\Phi\Phi}^\Phi
(\tilde{g}(t),t)\tilde{g}(t)^2\tfrac{m}{m+1}
\end{equation} 
higher order terms appear via the
off-critical structure constant $\tilde{C}_{\Phi\Phi}^\Phi(\tilde{g}(t),t)$. 
The structure constant $\tilde{C}^\Phi_{\Phi\Phi}(\tilde{g}(t),t)$ is
regular in the coupling $\tilde{g}$
\cite{guida,mikhak,cptzam}. 
The wilsonian effective action \eqref{SG2} is for zero coupling equal
to the Coulomb gas representation \eqref{actionSg} 
of the minimal model ${\cal M}_m$. 
The structure constant therefore has the following
expansion $\tilde{C}^\Phi_{\Phi\Phi}(\tilde{g}(t),t)
=\tilde{C}^\Phi_{\Phi\Phi}+O(\tilde{g}(t))$, where
$\tilde{C}^\Phi_{\Phi\Phi}$ is the structure constant for the vertex
operators $V_{(1,3)}$ in the minimal model ${\cal M}_m$.
$V_{(1,3)}(1)V_{(1,3)}(0)=\boldsymbol{1}+
\tilde{C}^\Phi_{\Phi\Phi}V_{(1,3)}(0)+\cdots$
and the structure constant for $\Phi_{(1,3)}$ is thus 
$C_{\Phi\Phi}^\Phi=N_{(1,3)}^{-1}\tilde{C}_{\Phi\Phi}^\Phi$
\cite{dotsenko}
\footnote{$(C_{\Phi\Phi}^{\Phi})^2\!=\!
\tfrac{16}{3}\tfrac{(1-y)^4}{(1-y/2)^2(1-3y/4)^2}
(\tfrac{\Gamma (1+y/2)}{\Gamma (1-y/2)})^4
(\tfrac{\Gamma (1-y/4)}{\Gamma (1+y/4)})^3
(\tfrac{\Gamma (1+y)}{\Gamma (1-y)})^2
(\tfrac{\Gamma (1+3y/4)}{\Gamma (1-3y/4)})$,
$C_{\Phi\Phi}^{\Phi}>0$. $\tilde{C}_{\Phi\Phi}^\Phi(\tilde{g}(t),t)$
can in principle be calculated in
conformal perturbation theory \cite{cptzam,guida,mikhak} 
for strictly relevant perturbations ($y>0$), and for small $t$ 
is it given by \cite{cptzam}
\begin{equation}
\tilde{C}_{\Phi\Phi}^\Phi (\tilde{g})= 
\tilde{C}_{\Phi\Phi}^\Phi -\tilde{g}\pi \int' d^2z\langle
V_{(1,3)}(\infty)V_{(1,3)}(z)V_{(1,3)}(1)V_{(1,3)}(0) \rangle_{{\cal M}_m}
+{\cal O}(\tilde{g}^3).
\end{equation}
The four-point function is the first non-trivial correlator and it can
be evaluated by pairing the fields and inserting the short distance
expansion. 
Unfortunately no closed form has been found for the conformal blocks
 with two screening charges ($\langle VVVV\rangle$
 requires the screening $Q_-^2$). When only one screening $Q_-$ is
 needed a closed expression of the conformal blocks can be found in
 terms of hyper-geometric functions using Euler's integral
 representation. In the marginal case $y=0$ the four-point
 function can be written in a closed form as the conformal blocks become
 meromorphic functions of $z$ \cite{laessig}, but conformal
 perturbation theory is only defined for strictly relevant 
 perturbations with $y>0$.}.

It follows from \eqref{sidsteeq} by inserting 
$\tilde{C}^\Phi_{\Phi\Phi}(\tilde{g}(t),t)
=\tilde{C}^\Phi_{\Phi\Phi}+O(\tilde{g}(t))$ that the infra-red fixed
point coupling vanishes in the limit 
$m\rightarrow \infty$, and this limit can therefore be compared with
the perturbative renormalisation group equation. For large $m$
\eqref{sidsteeq}
becomes 
$\dot{g}=yg-\pi C^\Phi_{\Phi\Phi}g^2+O(g^3)$ where
$g=N_{(1,3)}\tilde{g}$, and this has the infra-red fixed
point $g_{IR}=\frac{y}{\pi C^\Phi_{\Phi\Phi}}$ 
as obtained in \cite{zamorig,cardyludwig}. 

\section{Conclusion} 
We studied the exact renormalisation group equation for a two
dimensional quantum field theory.
It was approximated by only including
relevant operators of exponential form together with a background 
charge at infinity in the wilsonian effective action.
The effective action does not contain any derivatives in the field and
the approximation is therefore similar to the local potential
approximation, the non linear term in the exact renormalisation group
equation is approximated by the operator product expansion. 
We showed that
the effect of the background charge can be incorporated into the exact
renormalisation group equation by evaluating it at a point of non zero
curvature.

Using the equivalence between the unitary minimal models
perturbed by $\Phi_{(1,3)}$ and the quantum group restricted
sine-Gordon model, the obtained renormalisation group equation was
argued to describe the renormalisation group flow for the perturbed
unitary minimal models from ${\cal M}_m$ to ${\cal M}_{m-1}$. 
The resulting renormalisation group equation is valid to all orders in
the coupling for our truncation of the operator space 
(and for all $m>3$). The higher order terms in the coupling appear in 
the off-critical structure constant. 
In the limit of large $m$, where the ultra-violet
and infra-red fixed points approach each other, the renormalisation
group equation agrees with the well known perturbative result.

\ \\       

\acknowledgments
I thank the Danish Research Academy for a  research grant. I would
like to thank Paul Mansfield for reading the manuscript, and Vladimir
Dotsenko for sending me notes on the Coulomb gas representation.

\end{document}